%% file: ArxivVersion_Oct2011_jpg.tex
\documentclass[aps,prl,twocolumn,showpacs,superscriptaddress,groupedaddress]{revtex4}  
\usepackage{graphicx}  
\usepackage{dcolumn}   
\usepackage{bm}        
\usepackage{amssymb}   

\begin{document}

\title{Three-Axis Measurement and Cancellation of Background Magnetic Fields to less than 50 $\mu$G in a Cold Atom Experiment}
\input author_list.tex

\date{\today}

\begin{abstract}
Many experiments involving cold and ultracold atomic gases require very precise control of magnetic fields that couple to and drive the atomic spins.  Examples include quantum control of atomic spins, quantum control and quantum simulation in optical lattices, and studies of spinor Bose condensates.  This makes accurate cancellation of the (generally time dependent) background magnetic field a critical factor in such experiments. We describe a technique that uses the atomic spins themselves to measure DC and AC components of the background field independently along three orthogonal axes, with a resolution of a few tens of $\mu$G in a bandwidth of $\sim1$ kHz. Once measured, the background field can be cancelled with three pairs of compensating coils driven by arbitrary waveform generators. In our laboratory, the magnetic field environment is sufficiently stable for the procedure to reduce the field along each axis to less than $\sim$50 $\mu$G rms, corresponding to a suppression of the AC part by about one order of magnitude.  This suggests our approach can provide access to a new low-field regime in cold-atom experiments.
\end{abstract}

\maketitle

\section{\label{sec:level1} I.  Introduction}

Magnetic fields play an important role in the production, manipulation and study of cold and ultracold atomic gases.  A particular experiment may require a near-zero field environment or a very accurate applied field.  In either case, control of the total field requires a high level of background field suppression, and the degree to which this can be accomplished is often a key limitation.  One important example is quantum information processing and quantum simulation with qubits encoded in field sensitive atomic spin states.  Field sensitive states are required for spin-dependent atom transport in optical lattices \protect\cite{Mandel2003a}, and thus for studies of quantum walks \protect\cite{Karski2009} and the generation of entanglement via controlled collisions \protect\cite{Mandel2003b}\protect\cite{Anderlini2007}.  In such experiments, background magnetic fields limit the time and distance over which coherence and entanglement can be maintained.  Similarly, recent demonstrations of site resolved atom addressing in optical lattices use field dependent states \protect\cite{Lundblad2009}\protect\cite{Karski2010}\protect\cite{Weitenberg2011}, and background fields limit the spatial resolution and fidelity of control.  Going beyond qubits, the fidelity of quantum control and measurement of larger atomic spins is also fundamentally limited by background fields \protect\cite{Smith2006}\protect\cite{Chaudhury2007}\protect\cite{Chaudhury2009}.  A second important example is the study of spinor condensates \protect\cite{Pethick2002}, where many predictions have been made regarding novel ground states \protect\cite{Law1998}\protect\cite{Ho2000}, magnetic response \protect\cite{Koashi2000}, and dynamical control \protect\cite{Pu2000}\protect\cite{Duan2002} near zero magnetic field.  To reach this as yet inaccessible regime, background magnetic fields must be suppressed to a level where the Zeeman energy is negligible compared to the spin-dependent interaction energy.  In a typical experiment with $^{87}$Rb atoms in an optical dipole trap this may correspond to fields of a few tens of $\mu$G or less \protect\cite{Chang2004}.  Spinor condensates in much higher magnetic fields have been studied extensively \protect\cite{Stenger1998}\protect\cite{Schmaljohann2004}\protect\cite{Chang2004}\protect\cite{Chang2005}\protect\cite{Sadler2006}\protect\cite{Black2007}\protect\cite{Liu2009}, but even in this regime the ability to observe critical features of the dynamics can in some cases be affected by magnetic field stability.

Background magnetic fields are typically suppressed by passive magnetic shielding, or through measurement and active cancellation.  The latter is often preferable in experiments that require good optical access and/or rapidly time varying applied fields.  Conventional magnetometers cannot be placed at the position of the atom cloud, but it is possible in principle to interpolate the field from measurements with an array of sensors and cancel it in real-time using negative feedback \protect\cite{Ringot2001}.  Such schemes are limited by time varying field gradients and by the sensitivity and bandwidth of compact, affordable magnetometers, and cannot deliver the performance required for the most demanding experiments.  Alternatively, the experiment can be synchronized with the power line cycle, and contributions to the background field from DC and AC power-line components and from magnetization and eddy currents in the apparatus can be measured upfront and compensated with an applied field.  In an environment where these field components are dominant and stable, this approach can yield surprisingly good results.  In this paper we describe a novel technique whereby the cold atom sample itself is used as an $in$ $situ$ probe to measure the background magnetic field separately along each of three orthogonal axes, with a sensitivity of a few tens of $\mu$G in a bandwidth of $\sim$1 kHz. Three pairs of compensating coils driven by arbitrary waveform generators are subsequently used to cancel the measured field, and a second  measurement verifies that the residual field is below $\sim$50 $\mu$G rms.  Our scheme uses a spin echo technique to single out one of three orthogonal field components, and reads out the spin dynamics in real time by measuring the resulting polarization modulation of a weak, far-off-resonance optical probe.  We emphasize that the technique has been developed to meet a critical need in cold atom physics, rather than as a general purpose magnetometer.  General purpose magnetometry with hot \protect\cite{Budker2007}, cold \protect\cite{Terraciano2008}\protect\cite{Koschorreck2010}\protect\cite{Koschorreck2011} and ultracold \protect\cite{Vengalattore2007} atomic gasses has a considerable history, and it remains to be seen if the ideas outlined here might prove useful in that context.

The remainder of this paper is structured as follows.  In Sec. II we describe our experimental setup, including the optical readout of spin dynamics. Sections III and IV describe how we use spin echoes to measure and cancel field components that are transverse and longitudinal (parallel) to the probe propagation direction, respectively.  Section V describes how we measure and cancel the time dependent background field.  Finally, in Sec. VI we summarize our findings.

\section{\label{sec:level1} II.  Experimental Setup}

Our apparatus consists of a vapor cell magneto-optic trap (MOT)  and optical molasses, capable of preparing a sample of $\sim$10$^{7}$ Cs atoms at temperatures as low as 3 $\mu$K.  To reduce eddy currents and residual magnetization, we use an all glass vapor cell, our MOT coils are wound on a teflon support, and all magnetic or conductive materials are kept at a distance of at least 15 cm from the atom sample.  The vapor cell and MOT coils sit inside a precision machined plexiglass housing that holds three pairs of 7"$\times$7" square coils in near-Helmholz configurations.  These coil pairs generate magnetic fields along three precisely defined orthogonal directions, which we choose as the $x$, $y$, and $z$ axes of our coordinate system.  Driving each pair with an arbitrary waveform generator, we can apply fields up to 50 mG per axis, with a bandwidth of $\sim$350 kHz and an accuracy better than 1\%.  The primary use of these "control" coils is to generate time dependent fields for quantum control of atomic spins \protect\cite{Chaudhury2007}, but they are also the source of spin-echo pulses during field measurements, and AC compensation fields during subsequent experiments.  Cancellation of the DC  background field is performed with another, much larger set of "nulling" coils that surround the entire apparatus.  The experiment is triggered at a fixed point in the AC power cycle, so that the power-line contribution to the background field remains constant from one repetition to the next.  The size of our atom sample and its motion during the short duration of the experiment are small enough that spatial inhomogeneity of both the applied and background fields can be ignored.
	
	At the beginning of a measurement cycle we turn off the MOT coils and hold the atom sample in optical molasses for 15 ms, sufficient for the MOT field and associated eddy currents to decay completely.  The MOT/molasses beams are then extinguished and the atom sample is released into free fall.  At this point, we use a combination of optical pumping and Larmor precession in a pulsed magnetic field to initialize atoms in an $F = 3$ state with maximum projection of the spin along a desired direction.   An optical probe, initially polarized along \textit{x} and tuned in the vicinity of the Cs 6S$_{1/2}$ $\rightarrow$ 6P$_{1/2}$ (D1) transition, is passed through the atom sample along the \textit{z} (vertical) direction, and the resulting spin dependent change in its polarization state is measured with a low-noise polarimeter.  Because our atom sample has low-to-moderate optical density on resonance, $OD \leq 1$, the spin-probe coupling is too weak to generate significant entanglement, and its effects can be viewed as separate transformations of the spin and probe degrees of freedom.  Even so, for an atom with $F > 1/2$, the spin-probe interaction is both rich and complex (see ref. \protect\cite{Deutsch2010} for details) and we highlight only a few of the relevant features here.
	
	For a probe detuning much larger than the hyperfine splitting of the 6P$_{1/2}$ excited state, the dominant effect is Faraday rotation of the probe polarization. We determine the polarization rotation by measuring the power difference between the linear polarization components along  $(\bm{x}\pm\bm{y})/\sqrt{2}$, and obtain a signal 
\begin{equation}
\ M_{\rm Far}(t) \varpropto \frac{OD}{\Delta/\Gamma}\langle F_{z} \rangle_{t}, \
\end{equation}
where $\Delta$ is the detuning and $\Gamma$ = 4.7 MHz is the natural linewidth.  This is the basis for our measurements of magnetic fields transverse to the probe axis.  Fields along the probe axis conserve $\langle F_{z} \rangle$ and must be accessed by measuring a different spin observable.  In principle this can be done with a second probe beam propagating along, e.g.,  $x$ and measuring $\langle F_{x} \rangle$, but for technical reasons this is inconvenient to do in our setup.  We instead tune our probe beam between the $F' = 3$ and $F' = 4$ manifolds of the 6P$_{1/2}$ state, where the rank-2 tensor component of the atomic polarizability is substantial.  In this situation the atom sample becomes birefringent, and the resulting ellipticity of the probe polarization reflects the spin state.  We determine the ellipticity by measuring the power difference between the components of circular polarization and obtain a signal
\begin{equation}
\ M_{\rm Biref}(t) \varpropto \frac{OD}{(\Delta/\Gamma)^{2}}\langle F_{x}F_{y} + F_{y}F_{x} \rangle_{t}. \
\end{equation}
This signal is sensitive to longitudinal magnetic fields, going through two full periods each time the spin Larmor precesses by 2$\pi$ around the probe axis.
	
	For our choice of (linear) probe polarization, the probe induces a spin-dependent light shift of the form 
\begin{equation}
\ H_{\rm A} = \beta^{(2)}\hbar\gamma_{\rm s}F^2_x, \
\end{equation}	
where $\gamma_{\rm s}\varpropto(\Delta/\Gamma)^{-2}$ is the rate of photon scattering per atom, and where $\beta^{(2)}$ is a parameter that depends on the rank-2 tensor polarizability and thus on the precise probe frequency and details of the atomic transition. This nonlinear Hamiltonian causes a periodic collapse and revival of the mean spin \protect\cite{Smith2004}, which is undesirable in the present context. In our setup we are able to perform Faraday measurements, Eq. (1) with sufficient signal-to-noise ratio at modest probe power and detunings up to 100GHz, where the timescale for the first nonlinear collapse is much longer than the total measurement time.  By contrast, the birefiringence signal, Eq. (2), and nonlinear spin Hamiltonian both depend on the rank-2 tensor polarizability and thus scale in the same way with probe power and detuning.  As a result, we have been unable to find working conditions for which the nonlinear collapse can be ignored.  In this situation it is necessary to model the entire spin dynamics including nonlinear effects, in order to understand how the latter affect our magnetic field measurements.  See Sec. IV for further details.

\begin{figure}
\includegraphics[scale=1.1]{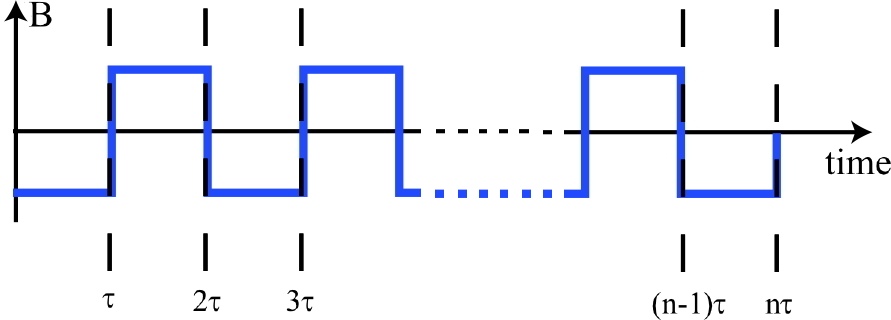}
\caption{\label{fig:epsart} Schematic of a rotary echo pulse sequence.  The magnitude of the applied magnetic field is constant in time but changes sign once per time interval $\tau$, generating a series of alternating $-2\pi$ and $+2\pi$ rotations around the field axis.  In our setup $B\approx50$ mG, corresponding to a Larmor period $\tau=57.0$ $\mu$s.}
\end{figure}

\section{\label{sec:level1} III.  Transverse Field Measurement}

We initiate a measurement of the background magnetic field along a transverse axis (e. g. the $x$ axis) by rotating the atomic spins so they are spin-up along $z$, $|\psi_{0}\rangle=|F=3,m_{z}=3\rangle$.  Following that, we apply a series of $n$ pulses of magnetic field along the measurement ($x$) axis, each having Larmor frequency $\omega_{\rm P}=2\pi\times17.5$ kHz ($B\approx50$ mG) and duration $\tau=2\pi/\omega_{\rm P}=57.0$ $\mu$s, and with the entire pulse train comprising a measurement window of duration $T=n\tau$.  The sign of the applied field is alternated from pulse to pulse (Fig. 1), so that the spins go through alternate rotations by $-2\pi$ and $+2\pi$ around the field axis.  This so-called $2\pi$ rotary spin echo sequence was originally developed by the NMR community \protect\cite{Vandersypen2005}, and has proven useful in other contexts including the manipulation of cold atom qubits \protect\cite{Rakreungdet2009}.  In our protocol it is key to isolating and measuring only the field component along the desired measurement axis.
	
\begin{figure}
\includegraphics[scale=1.22]{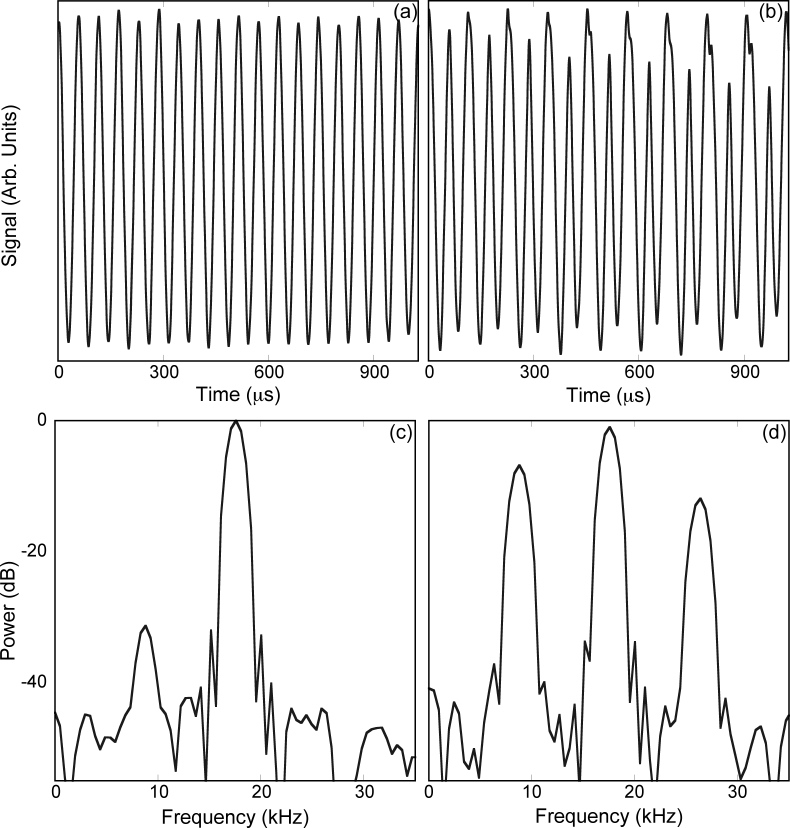}
\caption{\label{fig:epsart} Time-dependent polarization signals and corresponding power spectra in a transverse field measurement.  In the absence of a background field the signal from sequential rotations by $-2\pi$ and $+2\pi$ is indistinguishable from standard, unidirectional Larmor precession, resulting in a sinusoidal signal (a) and a power spectrum with a single peak at the Larmor frequency (c).  In the presence of a background field (280 $\mu$G in this example) kinks develop in the sinusoidal signal (b) and sidebands appear in the power spectrum (d).  The signals and power spectra shown here are averaged over 50 runs of the experiment.}
\end{figure}
	
	To understand how the rotary echo is used to generate a measurement signal, consider first the case of zero background field.  In this situation the direction of rotation is always reversed exactly when an atom returns to the spin-up state, and the Faraday signal $M_{\rm Far}(t)$ is a perfect sinusoid indistinguishable from Larmor precession in a constant field (Fig. 2(a)).  The power spectrum then consists of a single ÒcarrierÓ at frequency $1/\tau$ (Fig. 2(c)).  For a constant, non-zero background field along the measurement axis, the alternating pulse angles are $\pm2\pi+\omega_{\rm B}\tau$, where $\omega_{\rm B}$ is the background Larmor frequency, and successive reversals of rotation occur at points increasingly far from the spin-up state.  As a result, $M_{\rm Far}(t)$ develops a series of "kinks" (Fig. 2(b)), and those give rise to sidebands in the power spectrum that are shifted from the carrier frequency by $\pm1/2\tau$ (Fig. 2(d)).  A measurement of the background field can then be obtained by manually adjusting a compensating field to minimize the sidebands and null the total (average background plus compensating) field. 
	
\begin{figure}
\includegraphics[scale=1.21]{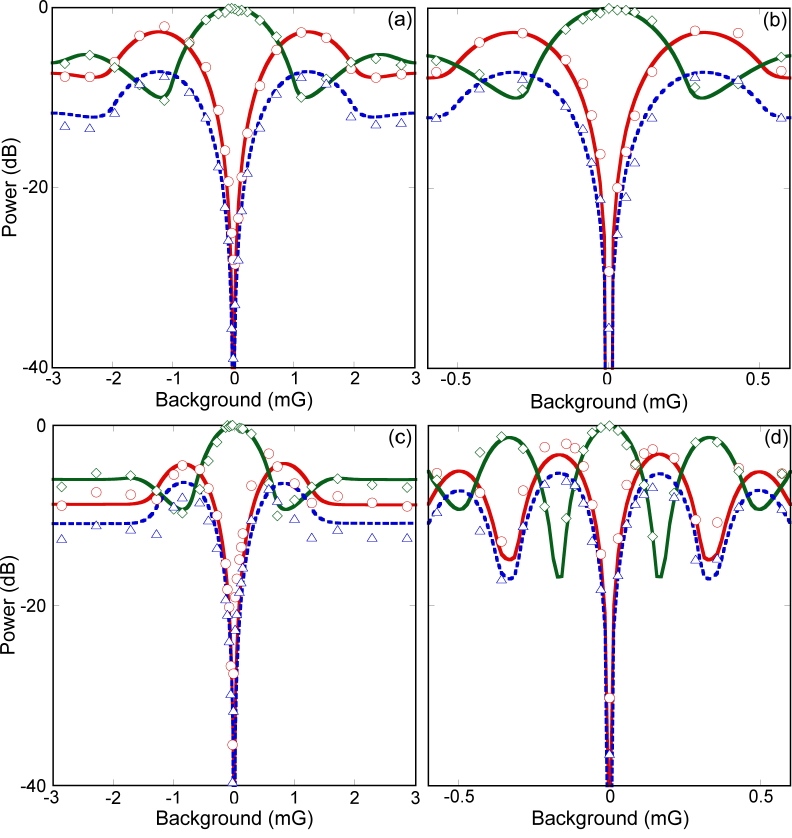}
\caption{\label{fig:epsart} (Color Online)  Power versus background field along the measurement axis, for the carrier (green/diamonds), and for the low-frequency (red/circles) and high-frequency (blue/triangles) sidebands.  Lines and symbols correspond to numerical simulation and experimental data, respectively.  In (a) and (b) the measurement axis is in the transverse direction, and the measurement windows are $T=1$ ms and $T=4$ ms respectively. In (c) and (d) the measurement axis is in the longitudinal direction, and the pulse durations are $T=1$ ms and $T=4$ ms respectively.  }
\end{figure}
	
	Figures 3(a) and 3(b) show calculated and measured powers in the carrier and sidebands as a function of a deliberately applied background field, and in particular the steep drop in the sideband powers near zero field.  Both these minima grow sharper with increasing measurement duration and/or increasing number of rotary echo pulses, leading to a tradeoff between measurement sensitivity and bandwidth as one would expect.  In the laboratory we typically average the power spectra from a few tens of repetitions, both to improve our signal to noise ratio and to reduce the sensitivity to field fluctuations that are not correlated with the AC power cycle.  Even so, with a cycle time near 1 second for our cold atom experiment, it is possible to perform the basic sideband minimization and field nulling routine in real time.  In practice, we are able to determine the point of minimum sideband power and thus measure the background field to within $\pm35$ $\mu$G for a measurement window  $T=1$ ms, and to within $\pm9$ $\mu$G for a measurement window $T=4$ ms.  

\begin{figure}
\includegraphics[scale=0.42]{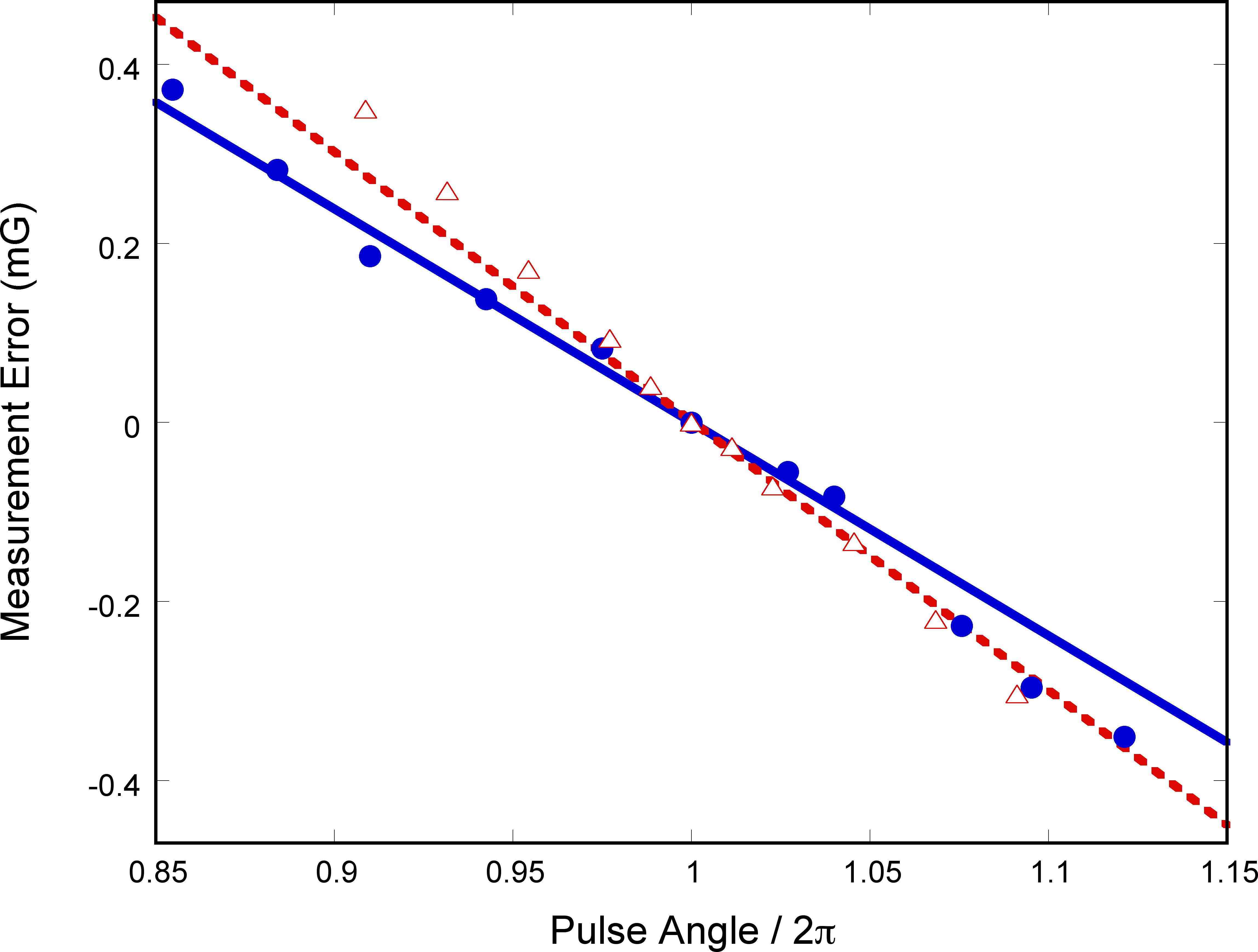}
\caption{\label{fig:epsart} (Color Online) Measurement error versus rotary echo pulse angle, for transverse (blue/circles) and longitudinal (red/triangles) field measurements.  Lines and symbols correspond to numerical simulation and experimental data, respectively.  As the pulse angle deviates from $2\pi$ the measured field is shifted from the true value. }
\end{figure}	
	
	It is important to consider how our measurement protocol is affected by errors in the magnetic field pulses that drive the rotary echo.  Asymmetry between positive and negative echo pulses mimic the presence of a background field, but can be easily detected by measuring the coil currents directly, and if necessary removed by reprogramming the arbitrary waveform generator.  Alternatively, the problem can be diagnosed by switching the leads to the relevant control coils between two consecutive measurements of the field, and looking for changes in  the apparent background field.  In our case, there is no indication that pulse asymmetry contributes significantly to the overall measurement uncertainty.  A second possibility is that the pulse areas, while equal, correspond to rotations by $\pm(2\pi+\epsilon)$, perhaps due to imperfect calibration of the control coils.  All rotary echoes are inherently robust against such angle errors -- the atom still returns to the spin-up state after every other pulse -- but the Faraday signal will have kinks and the power spectrum will contain sidebands even in the absence of a background magnetic field.  In spite of this, both numerical simulation and experiments with deliberately introduced angle errors show that minimization of the sideband power remains a good way to determine the background field.  Figure 4 shows the measurement error resulting from a given pulse angle error, defined as the shift in the required compensating field.  In our setup the estimated uncertainty in the pulse angle is $\pm0.5\%$, corresponding to a measurement error of $\pm13$ $\mu$G, which is less than our signal-to-noise limited resolution.
	
\begin{figure}
\includegraphics[scale=0.42]{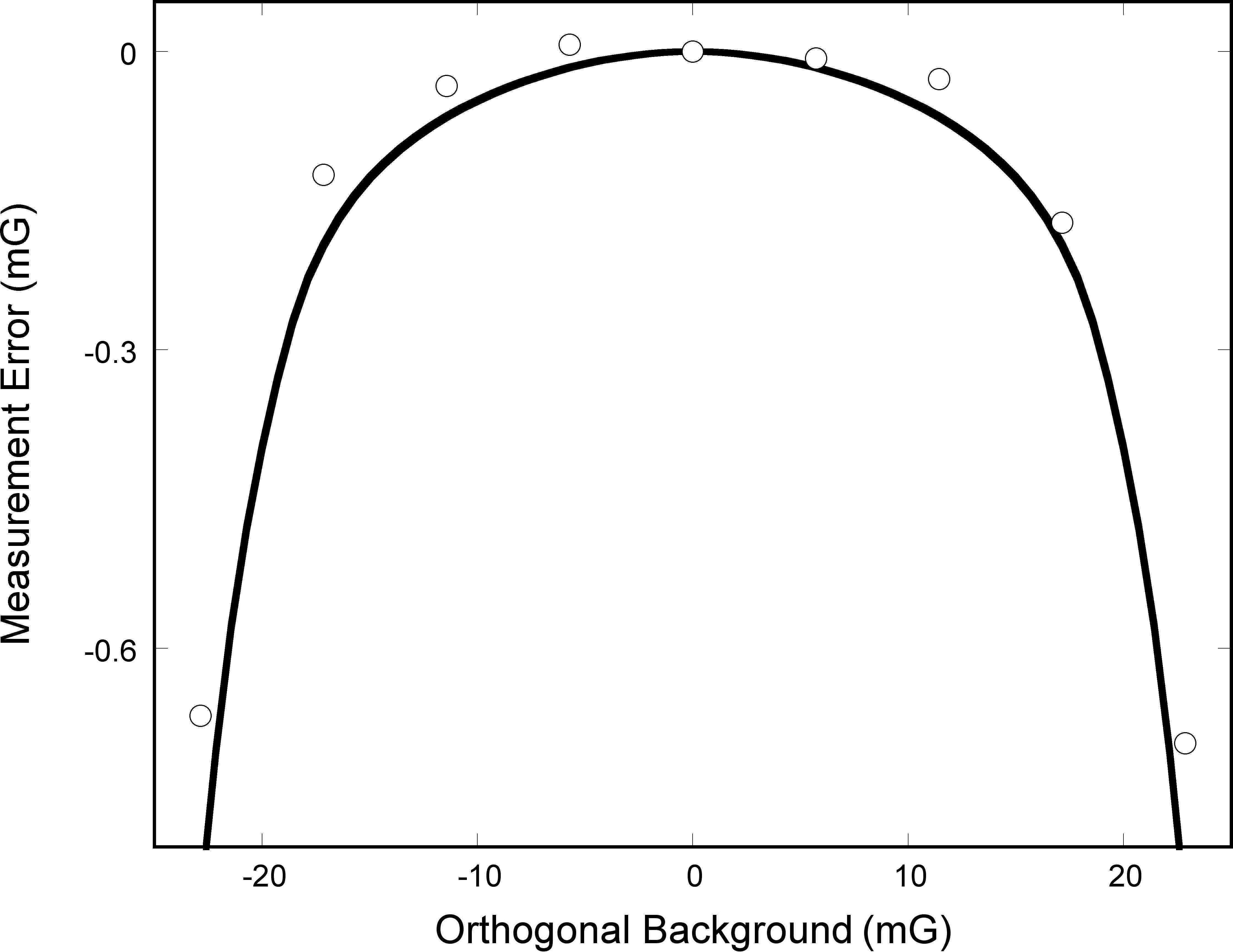}
\caption{\label{fig:epsart} Error in a transverse field measurement due to the presence of an orthogonal background field.  The lines and symbols correspond to numerical simulation and experimental data, respectively.}
\end{figure}
	
	It is known from NMR spectroscopy that $2\pi$ rotary echoes are robust not only against angle errors, but also against errors in the frequency of the driving field \protect\cite{Vandersypen2005}.  In our case, the equivalent of a frequency error is a non-zero field orthogonal to the measurement axis.  This suggests that our measurement should be close to uni-axial, unaffected by the presence of orthogonal fields.  Numerical simulations, as well as experimental measurements in the presence of deliberately applied orthogonal fields, confirm that the compensating field required to minimize sideband power shifts only very slightly, as shown in Fig. 5.  The resulting measurement error remains below 35 $\mu$G for orthogonal fields up to $\sim$8 mG, far beyond anything normally present in our apparatus.  To fully appreciate how small this effect is, consider an alternative scheme in which we apply a \textit{constant} rather than pulsed field along the measurement axis, and then measure the shift in the overall Larmor frequency $\omega$ resulting from a background field along the measurement axis.  In that case, the effect of orthogonal fields is also reduced because they add in quadrature with the bias field, $\omega=\sqrt{\omega^2_{\rm P}+\omega^2_\perp}$.  However, for our value of $\omega_{\rm P}$, a 8 mG orthogonal field would lead to a measurement error of 650 $\mu$G, a nearly twenty-fold increase relative to the rotary echo protocol!	
	
	Finally, it is worth noting that an orthogonal field large enough to produce a significant measurement error also leads to a large increase in sideband power even at the most optimal compensating field.  This serves as a convenient warning that a field measurement should not be trusted.  In this situation we resort to a simpler scheme, wherein we use our DC nulling coils to minimize Larmor precession in the absence of other applied fields.  This usually suffices to reduce the overall, time-averaged background field well below the 8 mG threshold.  Of course, once we have used our protocol to null the entire three dimensional background field, the orthogonal fields present in subsequent iterations of the measurement will be insignificant.

\section{\label{sec:level1} IV.  Longitudinal Field Measurement}

\begin{figure}
\includegraphics[scale=1.07]{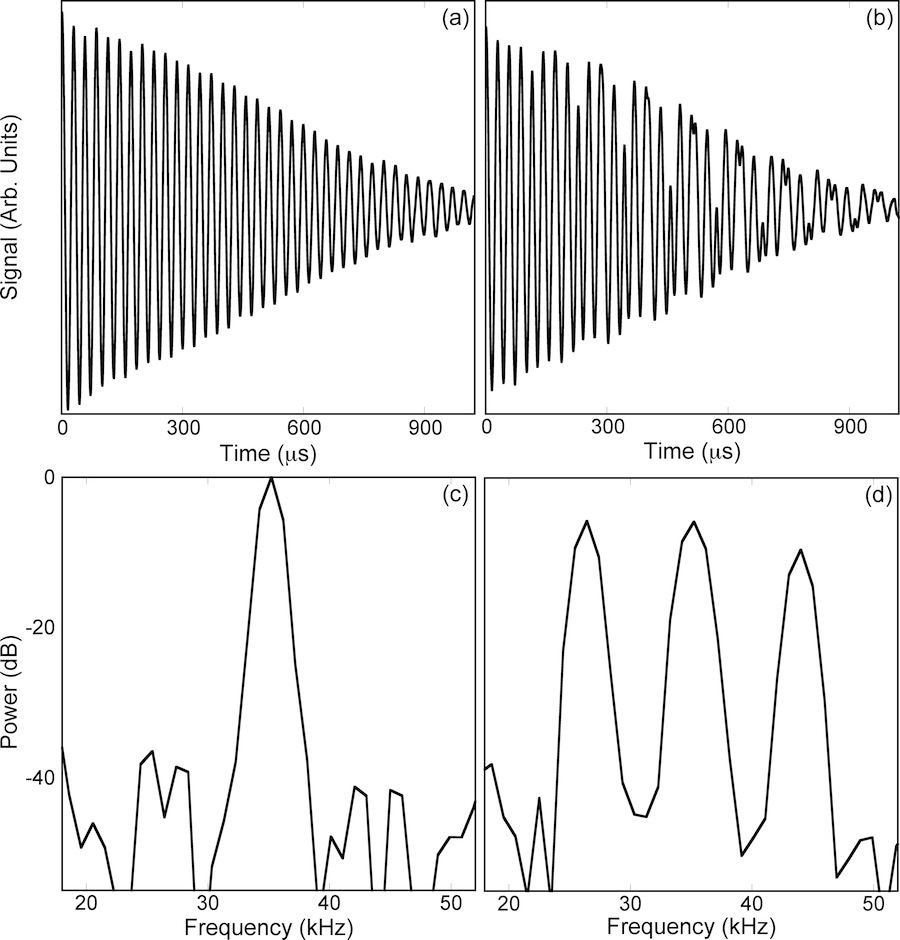}
\caption{\label{fig:epsart} Time dependent polarization signals and corresponding power spectra in a longitudinal field measurement.  In the absence of a background field the sequence of $\pm2\pi$ rotations produces a sinusoidal signal at twice the Larmor frequency (a), with a superimposed collapse due to the spin-dependent light shift. The corresponding power spectrum contains a single peak (c).  In the presence of a background field (430 $\mu$G in this example) kinks develop in the sinusoidal signal (b) and sidebands appear in the power spectrum (d).  The signals and power spectra shown here are averaged over 50 runs of the experiment.}
\end{figure}

As outlined in Sec. II, background fields parallel to the probe ($z$) axis must be measured using the birefringence signal, Eq. (2).  To initiate a measurement, we rotate the atomic spins so they are parallel to the $(\bm{x}+\bm{y})/\sqrt{2}$ axis and thus the expectation value $\langle F_xF_y+F_yF_x \rangle$ and $M_{\rm Biref}(t=0)$ take on their maximum value.  Following that, we apply the $2\pi$ rotary spin-echo sequence along the $z$ axis, and proceed just as for transverse fields.  Figures 6(a) and 6(b) show the birefringence signal for zero and non-zero background fields respectively, differing from the Faraday signals in transverse fields (Fig. 2) mainly by oscillating at twice the Larmor frequency and undergoing a nonlinear collapse in amplitude due to the spin-dependent light shift, Eq. (3).  The power spectra (Fig. 6(c) and Fig. 6(d)) consist of a carrier at frequency $2/\tau$, and sidebands shifted by $\pm1/2\tau$.  Longitudinal fields can now be measured in the same way as transverse fields, i. e. by applying a compensating field along the $z$ axis until the sideband power is minimized.  Figures 3(c) and (d) show the calculated and measured carrier and sideband powers versus background field, including minima near zero field of comparable width to those for transverse fields.
	
\begin{figure}
\includegraphics[scale=0.78]{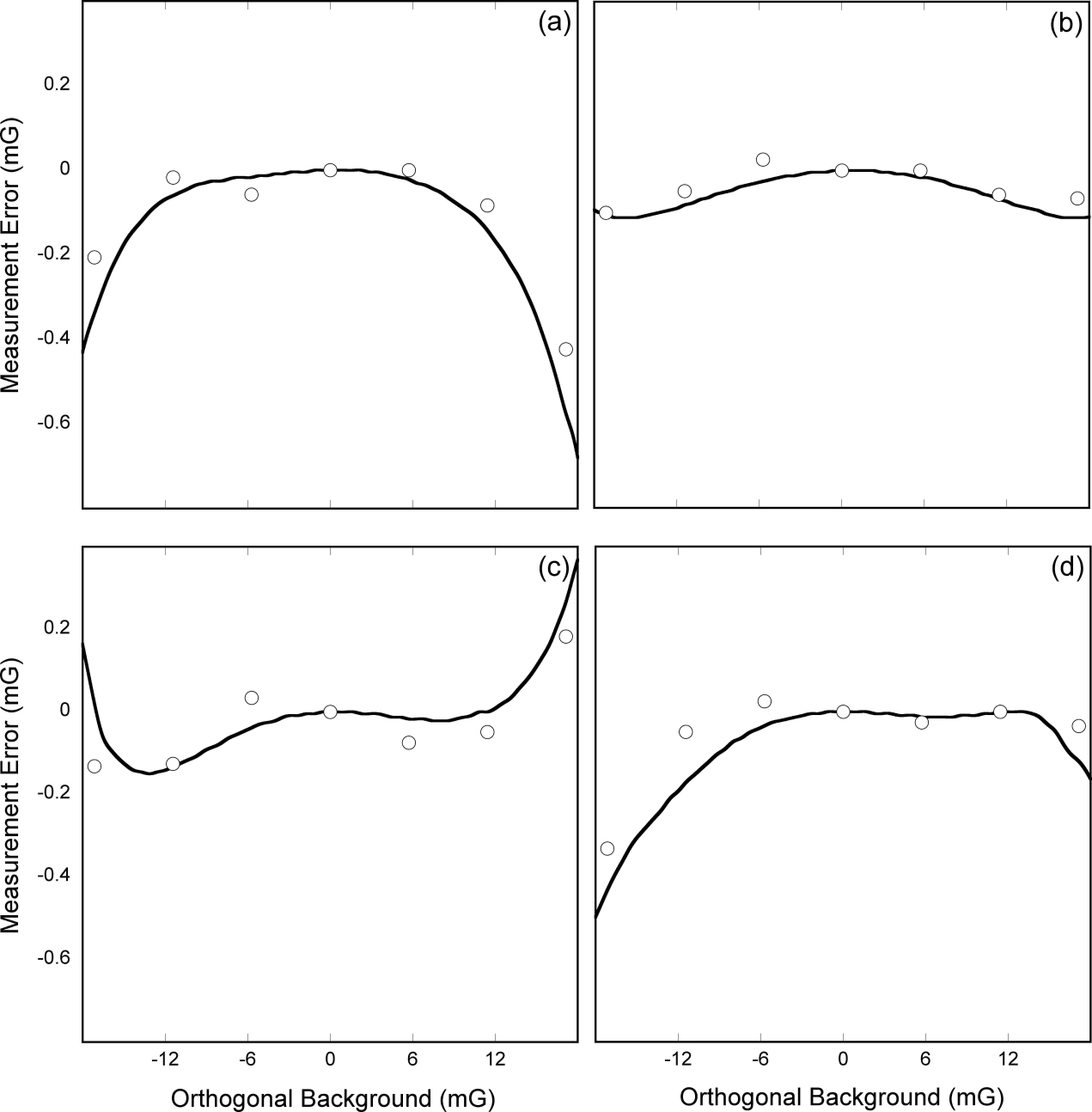}
\caption{\label{fig:epsart} Error in a longitudinal field measurement due to the presence of an orthogonal background field.  Lines and symbols correspond to numerical simulation and experimental data, respectively.  The orthogonal field is at an agle of $0^{\circ}$ (a), $45^{\circ}$ (b), $90^{\circ}$ (c) , and $135^{\circ}$ (d) with respect to the $x$ axis.}
\end{figure}	
	
The accuracy of longitudinal field measurements are subject to the same limitations as transverse field measurements.  Most importantly, the basic resolution limit, given by our ability to determine the point of minimum sideband power, is similar to that for transverse fields ($\pm35$ $\mu$G for $T=1$ ms,  $\pm9$ $\mu$G for $T=4$ ms).  Pulse angle errors and orthogonal background fields also play similar roles.  As shown in Fig. 4, the measurement error resulting from pulse angle errors is slightly larger, about $\pm15$ $\mu$G for our estimated $\pm0.5\%$ uncertainty in the pulse angle.  Figure 7 shows the measurement error as a function of orthogonal background fields.  Note that, in contrast to the case of transverse fields, the error for a longitudinal field depends on the \textit{direction} of the orthogonal field, presumably because both the signal $M_{\rm Biref}(t)$ and the non-linear light shift Hamiltonian $H_{\rm A}$ break the rotational symmetry in the $x$-$y$ plane.  Even so, we can still tolerate an orthogonal field as large as $\sim$5 mG before the corresponding error exceeds our 35 $\mu$G resolution limit.  As is the case for transverse fields, it is straightforward in practice to ensure that orthogonal fields are well below this limit.
	
\section{\label{sec:level1} V.  AC Background Field Measurement And Cancellation}

\begin{figure}
\includegraphics[scale=2.13]{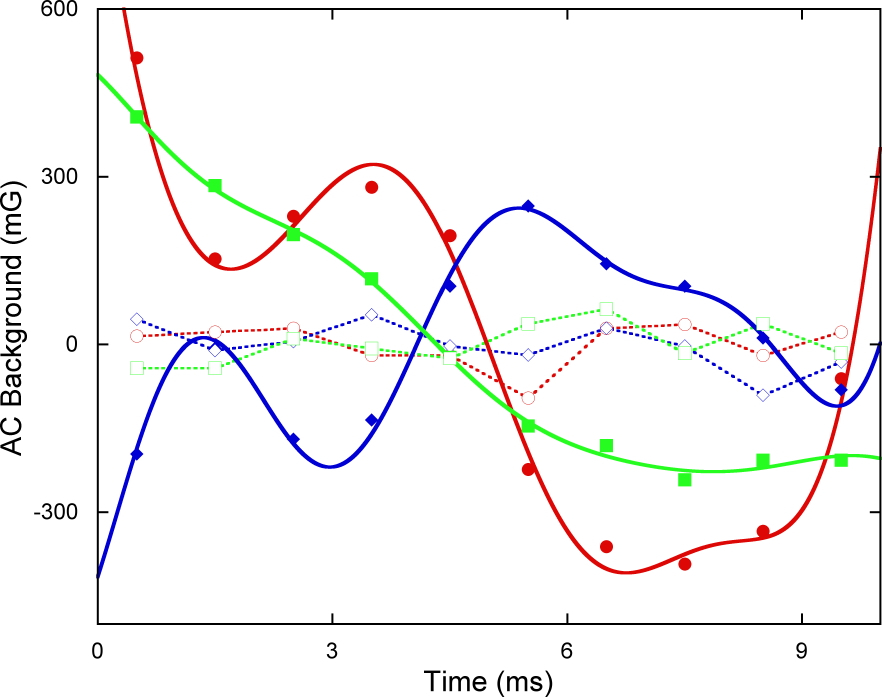}
\caption{\label{fig:epsart} (Color Online)  Example of AC power-line background fields in our laboratory, along directions $x$ (red), $y$ (blue) and $z$ (green) before (solid lines/solid symbols) and after (dashed lines/open symbols) cancellation.   Symbols indicate experimental data, solid lines are fits including DC and AC power line components.  The rms AC fields before (after) cancellation are 301 $\mu$G (38 $\mu$G) along $x$, 140 $\mu$G (39 $\mu$G) along $y$, and 223 $\mu$G (34 $\mu$G) along $z$.}
\end{figure}	

Our basic protocol for measuring background magnetic fields in short-time windows can be used to map out time-dependent fields with a time resolution given by the window width.  In practice, measuring the field for even a single time step requires a few runs of the experiment, so the approach is limited to fields that are stable and reproducible from run to run.  In our apparatus there is no significant contribution from residual magnetization or eddy currents, and the background field is dominated by DC and AC power-line components whose amplitude and phase are to a good approximation stable over time.  This can be easily verified with a three-axis fluxgate magnetometer placed near the vapor cell.  It is then straightforward to ensure reproducibility by triggering the start of each run at a fixed point in the 60 Hz power line cycle.  To map out the time dependent field in a given direction we perform a series of measurements spaced by the window width $T$.  We then fit this time series to a function containing DC, 60, 120, 180 and 240 Hz components, with the various amplitudes and phases as fit parameters.  To cancel the background field in subsequent experiments, we subtract the fit from whatever field is being applied by the arbitrary waveform generator and control coils for that axis.   Figure 8 shows an example of the measured three-axis AC field during a 10ms interval, before and after cancellation.   Empirically, we have found that the AC components of our background field tend to be stable over a period of many days, provided that we do not change the location or the on/off status of any electronic equipment in our lab.  In that case it suffices to re-measure and cancel them once a week.  The DC component is not as stable and must be re-measured and cancelled daily.  Overall, during periods when the magnetic field environment is quiet (nights or weekends) the cancelled background fields can remain below $\sim$50 $\mu$G rms per axis for several hours; similar but slightly worse performance can be achieved during standard working hours.

Finally, we note that time dependence can affect individual field measurements in non-intuitive ways.  For fields similar to those in Fig. 8 the time dependence can be considered roughly linear across any 1 ms measurement window.  For our parameters, numerical simulations show this will result in field estimates that are shifted from the mid-point value by 20$\%$ of the field change across the window, with a positive (negative) slope causing a negative (positive) offset.  As a result, the measured time-dependent fields turn out to be practically undistorted but shifted in time by $\sim200$ $\mu$s relative to the true fields.  For our modest AC fields this discrepancy is below the measurement accuracy, and we see no statistically significant difference in the cancelled fields whether we take the time shift into account or not.  Much larger AC fields might, however, require two or more iterations to achieve good cancellation.

\section{\label{sec:level1} VI. Summary and Outlook}

We have implemented and tested a protocol to perform $in$ $situ$ measurement and cancellation of DC and AC background magnetic fields in a cold atom experiment.  The protocol starts with a spin-polarized atom sample, applies a $2\pi$ rotary spin-echo sequence, and observes the resulting spin dynamics via polarization modulation of an optical probe.  A background field along the spin echo axis of rotation leads to distinctive sidebands in the power spectrum of the polarization modulation.  These sidebands can be minimized by canceling the background field with a known applied field, thereby yielding a measurement of the original background field.  Three-axis measurements can be performed by repeating the basic protocol with different spin echo axes of rotation, and time dependent fields can be mapped out by measuring the field in a series of short-time windows.  Once measured, a time-dependent field can be cancelled by adding a compensating field.  In this fashion, we routinely cancel our background field to below 50 $\mu$G rms per axis, limited by measurement resolution and the stability of the DC and AC power-line fields.   In our laboratory the approach has proven adequate for several demanding experiments involving non-trivial quantum control of atomic spins \protect\cite{Smith2006}\protect\cite{Chaudhury2007}\protect\cite{Chaudhury2009}.  Most recently, we have used a similar approach to measure and stabilize a dynamically switched 3 G magnetic bias field in a quantum control experiment now underway in our group \protect\cite{Merkel2008}.  Preliminary results indicate that, by canceling powerline components and compensating for transient effects following turn-on of the bias field, we are able to stabilize the total field (bias plus background) to within 30 $\mu$G  (one part in 10$^5$), and to set its value with an accuracy of 60 $\mu$G.  

We expect our magnetic field measurement and cancellation procedure will work at least as well in experiments with quantum degenerate gases.  In particular, the larger on-resonance optical density of these atomic samples should lead to much better signal-to-noise ratios in the polarimetry measurement \protect\cite{Smith2003}, and thus allow better measurement sensitivity and/or higher bandwidth.  Provided that such experiments are performed in magnetic field environments that are both quiet and stable, this may provide access to a new low-field regime for quantum simulation and spinor condensate physics.

This work was supported by ONR Grant No. N00014-05-1-420, and by NSF Grants PHY-0653631 and PHY-0903930.


\end{document}

%% file: author_list.tex
\affiliation{Center For Quantum Information and Control (CQuIC) and Department of Physics, University of Arizona, Tucson, Arizona 85721, USA}
\affiliation{Center For Quantum Information and Control (CQuIC) and College of Optical Sciences, University of Arizona, Tucson, Arizona 85721, USA}

\author{Aaron~Smith}\affiliation{Center For Quantum Information and Control (CQuIC) and Department of Physics, University of Arizona, Tucson, Arizona 85721, USA}
\author{Brian E.~Anderson} \affiliation{Center For Quantum Information and Control (CQuIC) and Department of Physics, University of Arizona, Tucson, Arizona 85721, USA}
\author{Souma~Chaudhury}\affiliation{Center For Quantum Information and Control (CQuIC) and College of Optical Sciences, University of Arizona, Tucson, Arizona 85721, USA}
\author{Poul S.~Jessen}\affiliation{Center For Quantum Information and Control (CQuIC) and Department of Physics, University of Arizona, Tucson, Arizona 85721, USA}\affiliation{Center For Quantum Information and Control (CQuIC) and College of Optical Sciences, University of Arizona, Tucson, Arizona 85721, USA}

\vskip 0.25cm

%% file: ArxivVersion_Oct2011_jpg.bbl
\begin{thebibliography}{21}
\expandafter\ifx\csname natexlab\endcsname\relax\def\natexlab#1{#1}\fi
\expandafter\ifx\csname bibnamefont\endcsname\relax
  \def\bibnamefont#1{#1}\fi
\expandafter\ifx\csname bibfnamefont\endcsname\relax
  \def\bibfnamefont#1{#1}\fi
\expandafter\ifx\csname citenamefont\endcsname\relax
  \def\citenamefont#1{#1}\fi
\expandafter\ifx\csname url\endcsname\relax
  \def\url#1{\texttt{#1}}\fi
\expandafter\ifx\csname urlprefix\endcsname\relax\def\urlprefix{URL }\fi
\providecommand{\bibinfo}[2]{#2}
\providecommand{\eprint}[2][]{\url{#2}}

\bibitem[{\citenamefont{{O. Mandel, M. Greiner, A. Widera, T. Rom, T. W. H\"ansch, and I. Bloch}}(2003}]{Mandel2003a}
\bibinfo{author}{\bibnamefont{{O. Mandel, M. Greiner, A. Widera, T. Rom, T. W. H\"ansch, and I. Bloch}}},
\bibinfo{journal}{Phys.\ Rev.\ Lett.} \textbf{\bibinfo{volume}{91}},
\bibinfo{pages}{010407} (\bibinfo{year}{2003}).

\bibitem[{\citenamefont{{M. Karski, L. F\"orster, J.-M. Choi, A. Steffen, W. Alt, D. Meschede, and A. Widera}}(2010)}]{Karski2009}
\bibinfo{author}{\bibnamefont{{M. Karski, L. F\"orster, J.-M. Choi, A. Steffen, W. Alt, D. Meschede, and A. Widera}}},
\bibinfo{journal}{Science} \textbf{\bibinfo{volume}{325}},
\bibinfo{pages}{174} (\bibinfo{year}{2009}).

\bibitem[{\citenamefont{{O. Mandel, M. Greiner, A. Widera, T. Rom, T. W. H\"ansch, and I. Bloch}}(2003}]{Mandel2003b}
\bibinfo{author}{\bibnamefont{{O. Mandel, M. Greiner, A. Widera, T. Rom, T. W. H\"ansch, and I. Bloch}}},
\bibinfo{journal}{Nature} \textbf{\bibinfo{volume}{425}},
\bibinfo{pages}{937} (\bibinfo{year}{2003}).

\bibitem[{\citenamefont{{M. Anderlini, P. J. Lee, B. L. Brown, J. Sebby-Trabley, W. D. Phillips, and J. V. Porto}}(2007}]{Anderlini2007}
\bibinfo{author}{\bibnamefont{{M. Anderlini, P. J. Lee, B. L. Brown, J. Sebby-Trabley, W. D. Phillips, and J. V. Porto}}},
\bibinfo{journal}{Nature} \textbf{\bibinfo{volume}{448}},
\bibinfo{pages}{452} (\bibinfo{year}{2007}).

\bibitem[{\citenamefont{{N. Lundblad, J. M. Obrecht, I. B. Spielman, and J. V. Porto}}(2009)}]{Lundblad2009}
\bibinfo{author}{\bibnamefont{{N. Lundblad, J. M. Obrecht, I. B. Spielman, and J. V. Porto}}},
\bibinfo{journal}{Nature\ Phys.} \textbf{\bibinfo{volume}{5}},
\bibinfo{pages}{575} (\bibinfo{year}{2009}).

\bibitem[{\citenamefont{{M. Karski, L. F\"orster, J.-M. Choi, A. Steffen, N. Belmechri, W. Alt, D. Meschede, and A. Widera}}(2010)}]{Karski2010}
\bibinfo{author}{\bibnamefont{{M. Karski, L. F\"orster, J.-M. Choi, A. Steffen, N. Belmechri, W. Alt, D. Meschede, and A. Widera}}},
\bibinfo{journal}{New\ J.\ Phys.} \textbf{\bibinfo{volume}{12}},
\bibinfo{pages}{065027} (\bibinfo{year}{2010}).

\bibitem[{\citenamefont{{C. Weitenberg, M. Endres, J. F. Sherson, M. Cheneau, P. Schauss, T. Fukuhara, I. Block, and S. Kuhr}}(2011)}]{Weitenberg2011}
\bibinfo{author}{\bibnamefont{{C. Weitenberg, M. Endres, J. F. Sherson, M. Cheneau, P. Schauss, T. Fukuhara, I. Block, and S. Kuhr}}},
\bibinfo{journal}{Nature} \textbf{\bibinfo{volume}{471}},
\bibinfo{pages}{319} (\bibinfo{year}{2011}).

\bibitem[{\citenamefont{{G. A. Smith, A. Silberfarb, I. H. Deutsch, and P. S. Jessen}}(2006}]{Smith2006}
\bibinfo{author}{\bibnamefont{{G. A. Smith, A. Silberfarb, I. H. Deutsch, and P. S. Jessen}}},
\bibinfo{journal}{Phys.\ Rev.\ Lett.} \textbf{\bibinfo{volume}{97}},
\bibinfo{pages}{180403} (\bibinfo{year}{2006}).

\bibitem[{\citenamefont{{S. Chaudhury, S. Merkel, T. Herr, A. Silberfarb, I. H. Deutsch, and P. S. Jessen}}(2007)}]{Chaudhury2007}
\bibinfo{author}{\bibnamefont{{S. Chaudhury, S. Merkel, T. Herr, A. Silberfarb, I. H. Deutsch, and P. S. Jessen}}},
\bibinfo{journal}{Phys.\ Rev.\ Lett.} \textbf{\bibinfo{volume}{99}},
\bibinfo{pages}{163002} (\bibinfo{year}{2007}).

\bibitem[{\citenamefont{{S. Chaudhury, A. Smith, B. E. Anderson, S. Ghose, and P. S. Jessen}}(2009)}]{Chaudhury2009}
\bibinfo{author}{\bibnamefont{{S. Chaudhury, A. Smith, B. E. Anderson, S. Ghose, and P. S. Jessen}}},
\bibinfo{journal}{Nature} \textbf{\bibinfo{volume}{461}},
\bibinfo{pages}{768} (\bibinfo{year}{2009}).

\bibitem[{\citenamefont{C. J. Pethick and H. Smith}(2002)}]{Pethick2002}
\bibinfo{author}{\bibnamefont{{See, for example, C. J. Pethick and H. Smith}}},
 \textit{\bibinfo{book}{Bose-Einstein Condensation in Dilute Gases}},(\bibinfo{publisher}{Cambridge University Press,
Cambridge, UK}, \bibinfo{year}{2002}).

\bibitem[{\citenamefont{{C. K. Law, H. Pu, and N. P. Bigelow}}(1998}]{Law1998}
\bibinfo{author}{\bibnamefont{{C. K. Law, H. Pu, and N. P. Bigelow}}},
\bibinfo{journal}{Phys.\ Rev.\ Lett.} \textbf{\bibinfo{volume}{81}},
\bibinfo{pages}{5257} (\bibinfo{year}{1998}).

\bibitem[{\citenamefont{{T.-L. Ho and S. K. Yip}}(2000}]{Ho2000}
\bibinfo{author}{\bibnamefont{{T.-L. Ho and S. K. Yip}}},
\bibinfo{journal}{Phys.\ Rev.\ Lett.} \textbf{\bibinfo{volume}{84}},
\bibinfo{pages}{4031} (\bibinfo{year}{2000}).

\bibitem[{\citenamefont{{M. Koashi and M. Ueda}}(2000}]{Koashi2000}
\bibinfo{author}{\bibnamefont{{M. Koashi and M. Ueda}}},
\bibinfo{journal}{Phys.\ Rev.\ Lett.} \textbf{\bibinfo{volume}{84}},
\bibinfo{pages}{1066} (\bibinfo{year}{2000}).

\bibitem[{\citenamefont{{H. Pu, S. Raghavan, and N. P. Bigelow}}(2000}]{Pu2000}
\bibinfo{author}{\bibnamefont{{H. Pu, S. Raghavan, and N. P. Bigelow}}},
\bibinfo{journal}{Phys.\ Rev.\ A} \textbf{\bibinfo{volume}{61}},
\bibinfo{pages}{023602} (\bibinfo{year}{2000}).

\bibitem[{\citenamefont{{L.-M. Duan, J. I. Cirac, and P. Zoller}}(2002}]{Duan2002}
\bibinfo{author}{\bibnamefont{{L.-M. Duan, J. I. Cirac, and P. Zoller}}},
\bibinfo{journal}{Phys.\ Rev.\ A} \textbf{\bibinfo{volume}{65}},
\bibinfo{pages}{033619} (\bibinfo{year}{2002}).

\bibitem[{\citenamefont{{M.-S. Chang, C. D. Hamley, M. D. Barrett, J. A. Sauer, K. M. Fortier, W. Zhang, L. You, and M. S. Chapman}}(2004}]{Chang2004}
\bibinfo{author}{\bibnamefont{{M.-S. Chang, C. D. Hamley, M. D. Barrett, J. A. Sauer, K. M. Fortier, W. Zhang, L. You, and M. S. Chapman}}},
\bibinfo{journal}{Phys.\ Rev.\ Lett.} \textbf{\bibinfo{volume}{92}},
\bibinfo{pages}{140403} (\bibinfo{year}{2004}).

\bibitem[{\citenamefont{{J. Stenger, S. Inouye, D. M. Stamper-Kurn, H.-J. Miesner, A. P. Chikkatur, and W. Ketterle}}(1998}]{Stenger1998}
\bibinfo{author}{\bibnamefont{{J. Stenger, S. Inouye, D. M. Stamper-Kurn, H.-J. Miesner, A. P. Chikkatur, and W. Ketterle}}},
\bibinfo{journal}{Nature} \textbf{\bibinfo{volume}{396}},
\bibinfo{pages}{345} (\bibinfo{year}{1198}).

\bibitem[{\citenamefont{{H. Schmaljohann, M. Erhard, J. Kronj\"ager, M. Kottke, S. Van Staa, L. Cacciapuoti, J. J. Arlt, K. Bongs, and K. Sengstock}}(2004}]{Schmaljohann2004}
\bibinfo{author}{\bibnamefont{{H. Schmaljohann, M. Erhard, J. Kronj\"ager, M. Kottke, S. van Staa, L. Cacciapuoti, J. J. Arlt, K. Bongs, and K. Sengstock}}},
\bibinfo{journal}{Phys.\ Rev.\ Lett.} \textbf{\bibinfo{volume}{92}},
\bibinfo{pages}{040402} (\bibinfo{year}{2004}).

\bibitem[{\citenamefont{{M.-S. Chang, Q. Quin, W. Zhang, L. You, and M. S. Chapman}}(2005}]{Chang2005}
\bibinfo{author}{\bibnamefont{{M.-S. Chang, Q. Quin, W. Zhang, L. You, and M. S. Chapman}}},
\bibinfo{journal}{Nature\ Phys.} \textbf{\bibinfo{volume}{1}},
\bibinfo{pages}{111} (\bibinfo{year}{2005}).

\bibitem[{\citenamefont{{L. E. Sadler, J. M. Higbie, S. R. Leslie, M. Vengalattore, and D. M. Stamper-Kurn}}(2006}]{Sadler2006}
\bibinfo{author}{\bibnamefont{{L. E. Sadler, J. M. Higbie, S. R. Leslie, M. Vengalattore, and D. M. Stamper-Kurn}}},
\bibinfo{journal}{Nature} \textbf{\bibinfo{volume}{443}},
\bibinfo{pages}{312} (\bibinfo{year}{2006}).

\bibitem[{\citenamefont{{A. T. Black, E. Gomez, L. D. Turner, S. Jung, and P. D. Lett}}(2007}]{Black2007}
\bibinfo{author}{\bibnamefont{{A. T. Black, E. Gomez, L. D. Turner, S. Jung, and P. D. Lett}}},
\bibinfo{journal}{Phys.\ Rev.\ Lett.} \textbf{\bibinfo{volume}{99}},
\bibinfo{pages}{070403} (\bibinfo{year}{2007}).

\bibitem[{\citenamefont{{Y. Liu, S. Jung, S. E. Maxwell, L. D. Turner, E. Tiesinga, and P. D. Lett}}(2009}]{Liu2009}
\bibinfo{author}{\bibnamefont{{Y. Liu, S. Jung, S. E. Maxwell, L. D. Turner, E. Tiesinga, and P. D. Lett}}},
\bibinfo{journal}{Phys.\ Rev.\ Lett.} \textbf{\bibinfo{volume}{102}},
\bibinfo{pages}{125301} (\bibinfo{year}{2009}).

\bibitem[{\citenamefont{{J. Ringot, P. Szriftgiser, and J. C. Garreau}}(2001}]{Ringot2001}
\bibinfo{author}{\bibnamefont{{J. Ringot, P. Szriftgiser, and J. C. Garreau}}},
\bibinfo{journal}{Phys.\ Rev.\ A} \textbf{\bibinfo{volume}{65}},
\bibinfo{pages}{013403} (\bibinfo{year}{2001}).

\bibitem[{\citenamefont{{D. Budker and M. Romalis}}(2007}]{Budker2007}
\bibinfo{author}{\bibnamefont{{See, for example, D. Budker and M. Romalis}}},
\bibinfo{journal}{Nature\ Phys.} \textbf{\bibinfo{volume}{3}},
\bibinfo{pages}{227} (\bibinfo{year}{2007}).

\bibitem[{\citenamefont{{M. L. terraciano, M. Bashkansky, and F. K. Fatemi}}(2008}]{Terraciano2008}
\bibinfo{author}{\bibnamefont{{M. L. Terraciano, M. Bashkansky, and F. K. Fatemi}}},
\bibinfo{journal}{Phys.\ Rev.\ A} \textbf{\bibinfo{volume}{77}},
\bibinfo{pages}{063417} (\bibinfo{year}{2008}).

\bibitem[{\citenamefont{{M. Koschorreck, M. Napolitano, B. Dubost, and M. W. Mitchell}}(2010}]{Koschorreck2010}
\bibinfo{author}{\bibnamefont{{M. Koschorreck, M. Napolitano, B. Dubost, and M. W. Mitchell}}},
\bibinfo{journal}{Phys.\ Rev.\ Lett.} \textbf{\bibinfo{volume}{104}},
\bibinfo{pages}{093602} (\bibinfo{year}{2010}).

\bibitem[{\citenamefont{{M. Koschorreck, M. Napolitano, B. Dubost, and M. W. Mitchell}}(2011}]{Koschorreck2011}
\bibinfo{author}{\bibnamefont{{M. Koschorreck, M. Napolitano, B. Dubost, and M. W. Mitchell}}},
\bibinfo{journal}{Appl.\ Phys.\ Lett.} \textbf{\bibinfo{volume}{98}},
\bibinfo{pages}{074101} (\bibinfo{year}{2011}).

\bibitem[{\citenamefont{{M. Vengalattore, J. M. Higbie, S. R. Leslie, J. Guzman, L. E. Sadler, and D. M. Stamper-Kurn}}(2007}]{Vengalattore2007}
\bibinfo{author}{\bibnamefont{{M. Vengalattore, J. M. Higbie, S. R. Leslie, J. Guzman, L. E. Sadler, and D. M. Stamper-Kurn}}},
\bibinfo{journal}{Phys.\ Rev.\ Lett.} \textbf{\bibinfo{volume}{98}},
\bibinfo{pages}{200801} (\bibinfo{year}{2007}).

\bibitem[{\citenamefont{{I. H. Deutsch and P. S. Jessen}}(2010}]{Deutsch2010}
\bibinfo{author}{\bibnamefont{{I. H. Deutsch and P. S. Jessen}}},
\bibinfo{journal}{Opt.\ Commun.} \textbf{\bibinfo{volume}{283}},
\bibinfo{pages}{681} (\bibinfo{year}{2010}).

\bibitem[{\citenamefont{{G. A. Smith, S. Chaudhury, A. Silberfarb, I. H. Deutsch, and P. S. Jessen}}(2004}]{Smith2004}
\bibinfo{author}{\bibnamefont{{G. A. Smith, S. Chaudhury, A. Silberfarb, I. H. Deutsch, and P. S. Jessen}}},
\bibinfo{journal}{Phys.\ Rev.\ Lett.} \textbf{\bibinfo{volume}{93}},
\bibinfo{pages}{163602} (\bibinfo{year}{2004}).

\bibitem[{\citenamefont{{L. M. K. Vandersypen and I. L. Chuang}}(2005}]{Vandersypen2005}
\bibinfo{author}{\bibnamefont{{L. M. K. Vandersypen and I. L. Chuang}}},
\bibinfo{journal}{Rev.\ Mod.\ Phys.} \textbf{\bibinfo{volume}{76}},
\bibinfo{pages}{1037} (\bibinfo{year}{2005}).

\bibitem[{\citenamefont{{W. Rakreungdet, J. H. Lee, K. F. Lee, B. E. Mischuck, E. Montano, and P. S. Jessen}}(2009}]{Rakreungdet2009}
\bibinfo{author}{\bibnamefont{{W. Rakreungdet, J. H. Lee, K. F. Lee, B. E. Mischuck, E. Montano, and P. S. Jessen}}},
\bibinfo{journal}{Phys.\ Rev.\ A} \textbf{\bibinfo{volume}{79}},
\bibinfo{pages}{022316} (\bibinfo{year}{2009}).

\bibitem[{\citenamefont{{S. T. Merkel, P. S. Jessen, and I. H. Deutsch}}(2008}]{Merkel2008}
\bibinfo{author}{\bibnamefont{{S. T. Merkel, P. S. Jessen, and I. H. Deutsch}}},
\bibinfo{journal}{Phys.\ Rev.\ A} \textbf{\bibinfo{volume}{78}},
\bibinfo{pages}{023404} (\bibinfo{year}{2008}).

\bibitem[{\citenamefont{{G. A. Smith, S. Chaudhury, and P. S. Jessen}}(2003}]{Smith2003}
\bibinfo{author}{\bibnamefont{{G. A. Smith, S. Chaudhury, and P. S. Jessen}}},
\bibinfo{journal}{J.\ Opt.\ B:\ Quantum\ Semiclass.\ Opt.} \textbf{\bibinfo{volume}{5}},
\bibinfo{pages}{323} (\bibinfo{year}{2003}).

\end{thebibliography}
